\documentclass[%
reprint,
superscriptaddress,
showpacs,preprintnumbers,
 amsmath,amssymb,
 aps,
longbibliography,
 prx,
floatfix,
]{revtex4-1}

\usepackage{float}
\usepackage{xspace}
\usepackage{graphicx}
\usepackage{dcolumn}
\usepackage{bm}
\usepackage{color}
\usepackage{soul}
\usepackage{multirow}
\usepackage{hyperref}
\hypersetup{
    unicode=false,          
    pdftoolbar=true,        
    pdfmenubar=true,        
    pdffitwindow=false,     
    pdfstartview={FitH},    
    pdftitle={Phonon anomalies in FeS},    
    pdfnewwindow=true,      
    colorlinks=true,       
    linkcolor=blue,          
    citecolor=blue,        
    filecolor=blue,      
    urlcolor=blue,       
    plainpages=false,        
}

\usepackage{CJK}

\newcommand{\CXT}{\mbox{CrXTe$_3$}\xspace}

\newcommand{\FGT}{\mbox{Fe$_{3-x}$GeTe$_2$}\xspace}

\newcommand{\Alg}{\texorpdfstring{\ensuremath{A_{1g}}\xspace}{A1g}}
\newcommand{\Elg}{\texorpdfstring{\ensuremath{E_{1g}}\xspace}{E1g}}
\newcommand{\EZg}{\texorpdfstring{\ensuremath{E_{2g}}\xspace}{E2g}}

\newcommand{\AZu}{\texorpdfstring{\ensuremath{A_{2u}}\xspace}{A2u}}
\newcommand{\Elu}{\texorpdfstring{\ensuremath{E_{1u}}\xspace}{E1u}}

\newcommand{\wn}{\ensuremath{\rm cm^{-1}}\xspace}
\begin{document}

\begin{CJK*}{GBK}{}

\title{Lattice dynamics and phase transitions in Fe$_{3-x}$GeTe$_2$}
\date{\today}
\author{A.~Milosavljevi\'{c}}
\affiliation{Center for Solid State Physics and New Materials, Institute of Physics Belgrade, University of Belgrade, Pregrevica 118, 11080 Belgrade, Serbia}
\author{A.~\v{S}olaji\'{c}}
\affiliation{Center for Solid State Physics and New Materials, Institute of Physics Belgrade, University of Belgrade, Pregrevica 118, 11080 Belgrade, Serbia}
\author{S. Djurdji\'{c}-Mijin}
\affiliation{Center for Solid State Physics and New Materials, Institute of Physics Belgrade, University of Belgrade, Pregrevica 118, 11080 Belgrade, Serbia}
\author{J.~Pe\v{s}i\'{c}}
\affiliation{Center for Solid State Physics and New Materials, Institute of Physics Belgrade, University of Belgrade, Pregrevica 118, 11080 Belgrade, Serbia}
\author{B. Vi\v{s}i\'{c}}
\affiliation{Center for Solid State Physics and New Materials, Institute of Physics Belgrade, University of Belgrade, Pregrevica 118, 11080 Belgrade, Serbia}
%
\author{Yu Liu} 
\affiliation{Condensed Matter Physics and Materials Science Department, Brookhaven National Laboratory, Upton, NY 11973-5000, USA}
\author{C.~Petrovic}
\affiliation{Condensed Matter Physics and Materials Science Department, Brookhaven National Laboratory, Upton, NY 11973-5000, USA}
\author{N.~Lazarevi\'{c}}
\affiliation{Center for Solid State Physics and New Materials, Institute of Physics Belgrade, University of Belgrade, Pregrevica 118, 11080 Belgrade, Serbia}
\author{Z.V.~Popovi\'{c}}
\affiliation{Center for Solid State Physics and New Materials, Institute of Physics Belgrade, University of Belgrade, Pregrevica 118, 11080 Belgrade, Serbia}
\affiliation{Serbian Academy of Sciences and Arts, Knez Mihailova 35, 11000 Belgrade, Serbia}

\begin{abstract}

We present Raman spectroscopy measurements of van der Waals bonded ferromagnet \FGT, together with  lattice dynamics. Four out of eight Raman active modes are observed and assigned, in agreement with numerical calculations. Energies and line-widths of the observed modes display unconventional temperature dependence at about 150\,K and 220\,K followed by the non-monotonic evolution of the Raman continuum. Whereas the former can be related to the magnetic phase transition, origin of the latter anomaly remains an open question.

\end{abstract}
\pacs{%
}
\maketitle

\end{CJK*}


\section{Introduction}

Novel class of magnetism hosting van der Waals bonded materials  have recently became of  great interest, since they are suitable candidates for numbers of technical applications \cite{PhysRevB.91.235425, Novoselov666, Wang2012, Gong2017, Huang2017}. Whereas \CXT (X = Si, Ge, Sn) and CrX$_3$ (X = Cl, Br, I) classes maintain low phase transitions temperatures
\cite{doi:10.1021/cm504242t, PhysRevB.91.235425, PhysRevB.92.035407, PhysRevB.95.245212, doi:10.1063/1.4914134, doi:10.1063/1.4914134} even in a monolayer regime \cite{C5TC03463A}, \FGT has a high bulk transition temperature, between 220\,K and 230\,K \cite{PhysRevB.93.144404, doi:10.7566/JPSJ.82.124711}, making it a promising applicant.

The \FGT crystal structure consists of Fe$_{3-x}$Ge sublayers stacked between two sheets of Te atoms, and a van der Waals gap between neighboring Te layers \cite{PhysRevB.96.144429,ejic.200501020}. Although structure contains two different types of Fe atoms, it is revealed that vacancies take  place only in the Fe2 sites \cite{PhysRevB.96.144429, PhysRevB.93.014411}.

Neutron diffraction, thermodynamic and transport measurements, and M\"{o}ssbauer spectroscopy were used to analyse magnetic and functional properties of \FGT, with Fe atoms deficiency of $x\approx0.1$ and $T_\mathrm{C} = 225$\,K. It is revealed that at temperature of 1.5\,K, magnetic moments of 1.95(5) and 1.56(4) $\mu_\mathrm{B}$ are directed along easy magnetic $c$-axes \cite{doi:10.1021/acs.inorgchem.5b01260}. In chemical vapor transport (CVT) grown Fe$_{3}$GeTe$_2$ single crystals, besides FM-PM transition at temperature of 214\,K, FM layers order antifferomagnetically at 152\,K \cite{Yi_2016}. Close to ferromagnetic transition temperature of 230\,K, possible Kondo lattice behaviour, i.e. coupling of travelling electrons and periodically localized spins is indicated at $T_\mathrm{K} = 190\pm20$\,K, which  is in a good agreement with theoretical predictions of 222\,K \cite{Zhangeaao6791}.

Lattice parameters, as well as magnetic transition temperature, vary with Fe ions concentration. Lattice parameters, $a$ and $c$ follow the opposite trend, whereas Curie temperature $T_\mathrm{C}$ decreases with an increase of Fe ions concentration\cite{PhysRevB.93.014411}. For flux-grown crystals, the critical behaviour was investigated by bulk \textit{dc} magnetization around the ferromagnetic phase  transition temperature  of 152\,K \cite{PhysRevB.96.144429}. Anomalous Hall effect was also studied, where a significant amount of defects produces bad metallic behaviour \cite{PhysRevB.97.165415}.

Theoretical calculations predict dynamical stability of Fe$_{3}$GeTe$_2$ single layer,  uniaxial magnetocrystalline anisotropy, that originates  from spin-orbit coupling \cite{PhysRevB.93.134407}. Recently, anomalous Hall effect measurements on single crystalline metallic Fe$_{3}$GeTe$_2$ nanoflakes with different thicknesses are reported, with a $T_\mathrm{C}$ near 200\,K and strong perpendicular magnetic anisotropy \cite{Tan2018}.

We report \FGT single crystal lattice dynamic calculations, together with Raman spectroscopy measurements. Four out of eight Raman active modes were observed and assigned. Phonon energies are in a good agreement with theoretical predictions. Analysed  phonon energies and line widths reveal fingerprint of ferromagnetic phase transition at temperature around 150\,K. Moreover, discontinuities in phonon properties are found at temperatures around 220\,K. Consistently, in the same temperature range, Raman continuum displays non-monotonic behaviour.

\section{Experiment and numerical method}
\label{sec:exp_theo}

\FGT single crystals were grown by self-flux method as previously described  \cite{PhysRevB.96.144429}. Samples for scanning electron microscopy (SEM) were cleaved and deposited on a graphite tape. Energy dispersive spectroscopy (EDS) maps were collected using FEI HeliosNanolab 650 equipped with an Oxford Instruments EDS system, equipped with an X-max SSD detector operating at 20 kV. The surface of the as-cleaved \FGT crystal appears to be uniform for several tens of microns in both directions, as shown in Fig.~\ref{fig:FigureSEM} of the Appendix.  Additionally, the elemental composition maps of Fe, Ge and Te show distinctive homogeneity of all the three elements [Fig.~\ref{fig:FigureEDS} of the Appendix].

For Raman scattering experiments, Tri Vista 557 spectrometer was used in the backscattering micro-Raman configuration. As an excitation source, solid state laser with 532 nm line was used. In our scattering configuration, plane of incidence is $ab$-plane, where $|a| = |b|$ ($\measuredangle (a,b) = 120^\circ$), with incident (scattered) light propagation direction along $c$-axes. Samples were cleaved in the air, right before being placed in the vacuum. All the measurements were performed in the high vacuum ($10^{-6}$ mbar) using a KONTI CryoVac continuous Helium flow cryostat with 0.5 mm thick window. To achieve laser beam focusing, microscope objective with $\times$50 magnification was used. Bose factor correction of all spectra was performed. More details can be found in the Appendix.
\begin{figure}[t!]
  \centering
  \includegraphics[width=85mm]{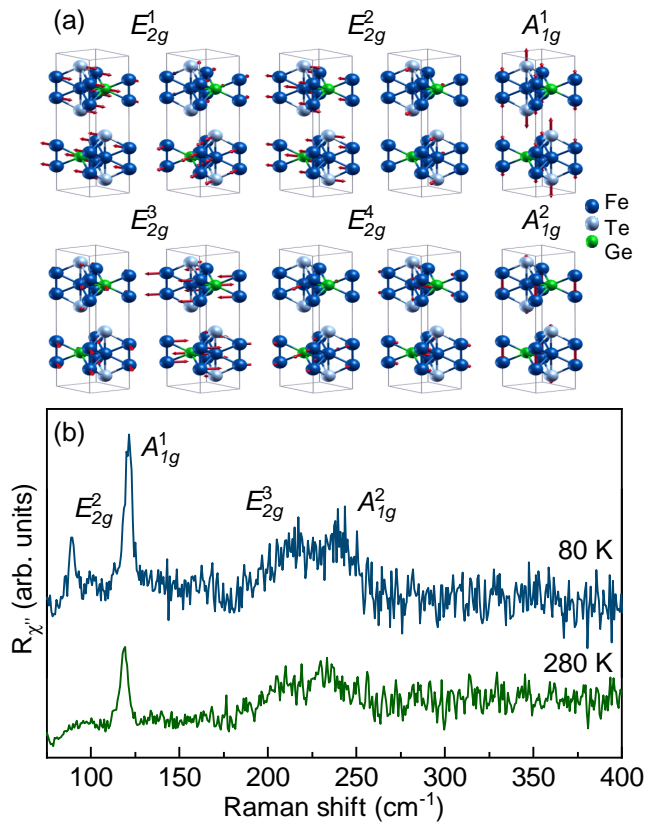}
  \caption{(Color  online) (a) Displacement patterns of \Alg and \EZg symmetry modes. (b) Raman spectra of \FGT single crystal measured at different temperatures in parallel polarization configuration.}
 \label{fig:Figure1}
\end{figure}

Density functional theory calculations were performed in Quantum Espresso software package \cite{QE-2009}. We used the PAW pseudopotentials \cite{PhysRevB.50.17953,PhysRevB.59.1758} with the Perdew-Burke-Ernzerhof (PBE) exchange-correlation functional \cite{PhysRevLett.77.3865}. The electron wavefunction and charge density cutoffs of 64 Ry and 782 Ry were chosen, respectively. The $k$-points were sampled using the Monkhorst-Pack scheme, with  $8\times8\times4$ $\Gamma$-centered grid. Both magnetic and non-magnetic calculations were performed, using the experimentally obtained lattice parameters and the calculated values obtained by relaxing the theoretically proposed structure. In order to obtain the lattice parameters accurately, treatment of the Van der Waals interactions is introduced. Van der Waals interaction was included in all calculations using the Grimme-D2 correction \cite{grimme2006semiempirical}. Phonon frequencies in $\Gamma$ point are calculated within the linear response method implemented in QE.

\section{Results and Discussion}
\label{sec:results}

\FGT crystallises in hexagonal crystal structure, described with $P6_3/mmc$ ($D_{6h}^4$) space group. The atom type, site symmetry, each site's contribution to the phonons in $\Gamma$ point, and corresponding Raman tensors for $P6_3/mmc$ space group are presented in Table~\ref{ref:Table1}.

\begin{table}
\caption{Top panel: The type of atoms, Wyckoff positions, each site's contribution to the phonons in $\Gamma$ point and corresponding Raman tensors for $P6_3/mmc$ space group of \FGT. Bottom panel: Phonon symmetry, calculated optical Raman active phonon frequencies (in \wn) for magnetic (M) phase, and experimental values for Raman active phonons at 80\,K.}
\label{ref:Table1}
\begin{ruledtabular}
\centering
\resizebox{\linewidth}{!}{%
\begin{tabular}{c c c}
\multicolumn{3}{c} {Space group  $P6_3/mmc$ (No. 194)} \\

\cline{1-3} \\[-0.5em]

Fe1 ($4e$)
& \multicolumn{2}{c}{$A_{1g} + E_{1g} + E_{2g}$+$A_{2u} + E_{1u}$} \\[1mm]

Fe2 ($2c$)
& \multicolumn{2}{c}{$E_{2g}$+$A_{2u} + E_{1u}$} \\[1mm]

Ge ($2d$)
& \multicolumn{2}{c}{$E_{2g}$+$A_{2u} + E_{1u}$} \\[1mm]

Te ($2c$)
& \multicolumn{2}{c}{$A_{1g} + E_{1g} + E_{2g}$+$A_{2u} + E_{1u}$} \\[1mm]
\hline \\[-0.5em]

\multicolumn{3}{c}{Raman tensors} \\[1mm] \hline \\[-0.5em]

$
A_{1g} = \begin{pmatrix}
a&0&0\\
0&a&0\\
0&0&b\\
\end{pmatrix}
$
&
$E_{1g} = \begin{pmatrix}
0&0&-c\\
0&0&c\\
-c&c&0\\
\end{pmatrix}
$
&
$E_{2g} = \begin{pmatrix}
d&-d&0\\
-d&-d&0\\
0&0&0\\
\end{pmatrix}$ \\[5mm] \cline{1-3}  \\[-0.5em]

\multicolumn{3}{c}{Raman active modes} \\ [1mm] \cline{1-3}\\[-0.3em]

Symmetry & Calculations (M) & Experiment (M)\\[1mm] \cline{1-3} \\[-0.5em]

$\EZg^1$ & 50.2  & - \\[1mm]
$\Elg^1$ & 70.3  & - \\[1mm]
$\EZg^2$ & 122.2  & 89.2  \\[1mm]
$\Alg^1$ & 137.2 & 121.1 \\[1mm]
$\Elg^2$ & 209.5 &  -  \\[1mm]
$\EZg^3$ & 228.6 & 214.8 \\[1mm]
$\Alg^2$ & 233.4 & 239.6  \\[1mm]
$\EZg^4$ & 334.3 & - \\[1mm]
\end{tabular}}
\end{ruledtabular}
\end{table}

Calculated displacement patterns of Raman active modes, which can be observed in our scattering configuration are  presented in Fig.~\ref{fig:Figure1}\,(a). Since Raman tensor of \Elg mode contains only $z$ component [Tab. \ref{ref:Table1}], by selection rules, it can not be detected  when measuring from the $ab$ plane in the backscattering configuration.
Whereas \Alg modes include vibrations of Fe and Te ions along $c$-axis, \EZg modes include in plane vibrations of all four atoms. Raman spectra of \FGT in magnetic phase (M), at 80\,K and non-magnetic phase (NM), at 280\,K, in parallel scattering configuration ($\mathbf{e}_i\parallel\mathbf{e}_s$), are presented in Fig.~\ref{fig:Figure1}\,(b). As it can be seen, four peaks at  89.2 \wn, 121.1 \wn, 214.8 \wn and 239.6 \wn can be clearly observed at 80\,K. According to numerical calculations [see Table~\ref{ref:Table1}], peaks at 89.2 \wn and 239.6 \wn correspond to two out of four \EZg modes, whereas peaks at 121.1 \wn and 239.6 \wn can be assigned as two \Alg symmetry modes. One should note that numerical calculations performed by using experimentally obtained lattice parameters in magnetic phase yield better agreement with experimental values. This is not surprising since the calculations are preformed for the stoichiometric compound as opposed to the non-stoichiometry of the sample. Furthermore, it is known that lattice parameters strongly depend on Fe atoms deficiency \cite{PhysRevB.93.014411}. All calculated Raman and infrared phonon frequencies, for magnetic and non magnetic phase of \FGT, using relaxed and experimental lattice parameters, together with experimentally observed Raman active modes are summarized in Table~\ref{ref:TableA1} of the Appendix.

\begin{figure}[t]
  \centering
  \includegraphics[width=85mm]{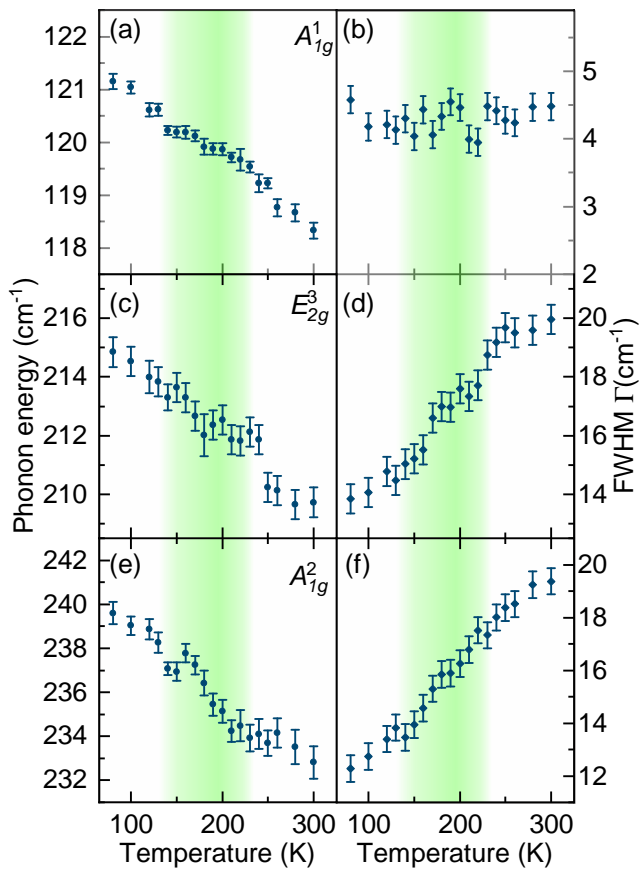}
  \caption{(Color  online) Energy  and linewidth  temperature dependence of $\Alg^1$ ((a) and (b)), $\EZg^3$ ((c) and (d)) and $\Alg^2$ ((e) and (f)) phonon modes in \FGT.}
 \label{fig:Figure2}
\end{figure}

After assigning all  observed modes we focused on their temperature evolution. Having in mind finite instrumental broadening, Voigt line shape was used for the data analysis\cite{Milosavljevic2018,Baum2018}.
Modelling procedure is described in details in the Appendix and presented in Fig.~\ref{fig:FigureA1}. Fig.~\ref{fig:Figure2} shows temperature evolution of energy and linewidth of $\Alg^1$, $\EZg^3$ and $\Alg^2$ mode between 80\,K and 300\,K. Upon heating the sample, both, energy and linewidth of $\Alg^1$ and $\Alg^2$ symmetry modes, exhibit small, but sudden discontinuity at about 150\,K [Fig.~\ref{fig:Figure2}\,(a) and (e)]. Apparent discontinuity in energy of all analysed Raman modes is again present at temperatures around 220\,K. In the same temperature range linewidths of these Raman modes show clear deviation from the standard anharmonic behavior \cite{Milosavljevic2018,Baum2018, Opacic2016, POPOVIC201451,PhysRevB.91.064303}.

%

\begin{figure}[t]
  \centering
  \includegraphics[width=85mm]{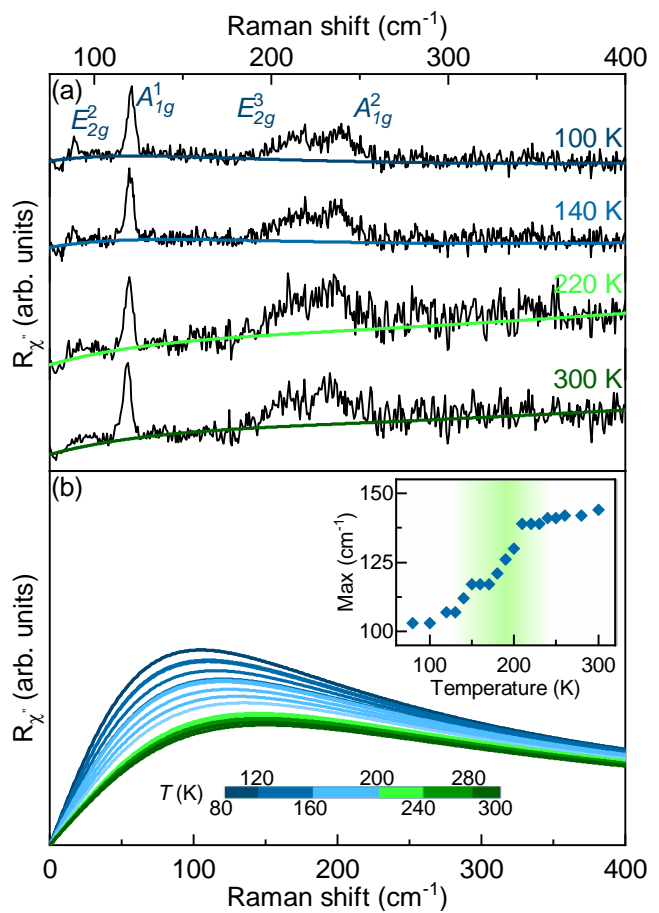}
  \caption{(Color  online) (a) Raman spectra of \FGT at four temperatures measured in parallel polarization configuration. Solid lines represent theoretical fit to the experimental data. (b) Temperature evolution of the electronic continuum after omitting linear therm. Inset: Displacement of the fitted curves maximum.}
 \label{fig:Figure3}
\end{figure}

Apart from the anomalies in the phonon spectra, closer inspection of the temperature dependent Raman spectra measured in the parallel polarization configuration reveals pronounced evolution of the Raman continuum [Fig.~\ref{fig:Figure3}\,(a)]. For the analysis we have used simple model including damped Lorentzian and linear therm, $\chi_{cont}'' \propto a \Gamma \omega/(\omega^2 + \Gamma^2) + b \omega$ \cite{Hackl2007}, where $a$, $b$ and $\Gamma$ are temperature dependent parameters. Fig.~\ref{fig:Figure3}\,(b) summarizes the results of the analysis with linear therm omitted (most likely originating from a luminescence). At approximately same temperatures, where phonon properties exhibit discontinuities, the continuum temperature dependence manifests non-monotonic behaviour. The curve maximum positions were obtained by integrating those shown in Fig.~\ref{fig:Figure3}\,(b). Inset of Fig.~\ref{fig:Figure3}\,(b) shows temperature evolution of their displacements. This analysis confirms the presence of discontinuities in electronic continuum at temperatures around 150\,K and 220\,K, which leaves a trace in phonon behaviour around these temperatures [Fig. \ref{fig:Figure2}]. While we do not have evidence for Kondo effect in \FGT crystals we measured, modification of electronic background at FM ordering due to localization or Kondo effect cannot be excluded.


The temperature evolutions of phonon self-energies and the continuum observed in the Raman spectra of \FGT, suggest the presence of phase transition(s). Magnetization measurements of samples were performed as described in Ref.~\cite{PhysRevB.96.144429}, revealing FM-PM transition at 150\,K. Thus, the discontinuity in observed phonon properties around this temperature, can be traced back to the weak to moderate spin-phonon coupling. The question remains open regarding the anomaly  observed at about 220\,K. As previously reported, Curie temperature of the \FGT single crystals grown by CVT method is between 220\,K and 230\,K \cite{ejic.200501020, PhysRevB.93.144404,doi:10.7566/JPSJ.82.124711}, varying with vacancies concentration, i.e. decrease in vacancies content will result the increment of $T_\mathrm{C}$ \cite{PhysRevB.93.014411}. On the other hand, the \FGT crystals grown by self-flux method usually have lower Curie temperature, since the vacancy content is higher \cite{PhysRevB.96.144429, PhysRevB.93.014411}. Crystals used in the Raman scattering experiment presented here were grown by self-flux method with the Fe vacancy content of $x \approx 0.36$ \cite{PhysRevB.96.144429}. This is in a good agreement with our EDS results of $x = 0.4 \pm 0.1$, giving rise to the FM-PM transition at 150\,K. Nevertheless, inhomogeneous distribution of vacancies may result in a formation of vacancy depleted ''islands'' which in turn would result in anomaly at 220\,K similar to the one observed in our Raman data. However, the EDS data [see Fig.~\ref{fig:FigureEDS}] do not support this possibility.
At this point we can only speculate that while the long-range order temperature is shifted to lower temperature by introduction of vacancies, short range correlations may develop at 220\,K.


\section{Conclusion}
\label{sec:conclusion}

We have studied lattice dynamic of flux-grown \FGT single crystals by means of Raman spectroscopy and DFT. Four out of eight Raman active modes, two \Alg and two \EZg, have been observed and assigned. DFT calculations are in a good agreement with experimental results. Temperature dependence of $\Alg^1$, $\EZg^3$ and $\Alg^2$ mode properties reveals clear fingerprint of spin-phonon coupling, at temperature around 150\,K. Furthermore, anomalous behaviour in energies and line widths of observed phonon modes is present in Raman spectra at temperatures around 220\,K with the discontinuity also present in the electronic continuum. Its origin still  remains an open question, and requests further analysis.

\section*{Acknowledgement}
The work was supported by the Serbian Ministry of Education, Science and Technological Development under Projects III45018 and OI171005. DFT calculations were performed using computational resources at Johannes Kepler University, Linz, Austria.  Materials synthesis was supported by the U.S. Department of Energy, Office of Basic Energy Sciences as part of the Computation Material Science Program (Y.L. and C.P.). Electron microscopy was performed at Jozef Stefan Institute, Ljubljana, Slovenia under Slovenian Research Agency contract P1-0099 (B. V.).


%



\clearpage
\begin{appendix}

\setcounter{figure}{0}
\renewcommand\thefigure{A\arabic{figure}}

\setcounter{table}{0}
\renewcommand\thetable{A\Roman{table}}

\section*{Appendix A: Electron microscopy} 
\label{sec:sem}

\begin{figure}[h!]
  \centering
  \includegraphics[width=85mm]{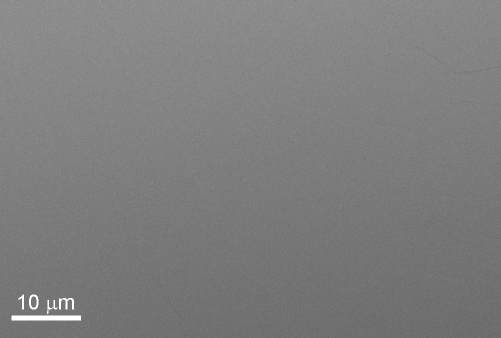}
  \caption{(Color  online) SEM image of a \FGT single crystal.}
 \label{fig:FigureSEM}
\end{figure}

\begin{figure}[bh!]
  \centering
  \includegraphics[width=85mm]{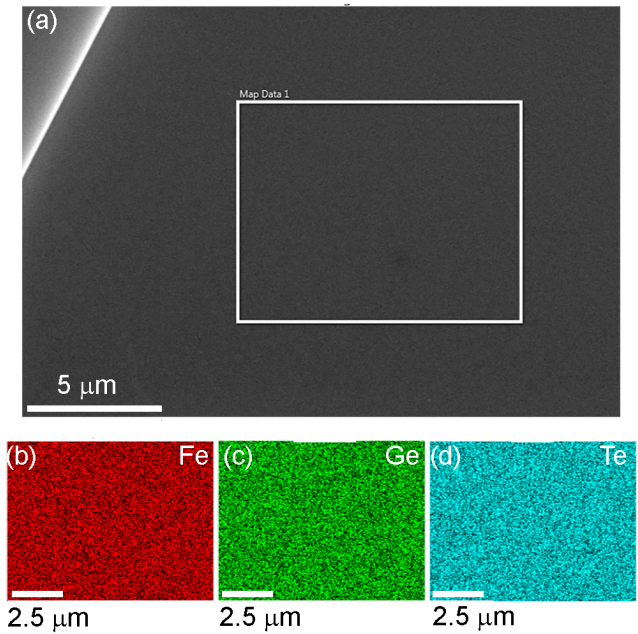}
  \caption{(Color  online) EDS mapping on a \FGT single crystal. (a) Secondary electron image of the crystal with the mapping performed within the rectangle. (b) - (d) Associated EDS maps for Fe, Ge and Te, respectively.}
 \label{fig:FigureEDS}
\end{figure}

In order to examine the uniformity of \FGT, Scanning electron microscopy was performed on as-cleaved crystals. It can be seen from Figure \ref{fig:FigureSEM} that the crystals maintain uniformity for several tens of microns. Furthermore, elemental composition was obtained using EDS mapping, as shown in \ref{fig:FigureEDS}. The atomic percentage, averaged over ten measurements, is 47, 17 and 36\% ($\pm 2 \%$) for Fe, Ge and Te, respectively, with the vacancy content $x = 0.4 \pm 0.1$. The maps associated with the selected elements appear homogeneous, as they are all present uniformly with no apparent islands or vacancies.

\section*{Appendix B: Data modelling}
\label{sec:Fit_res}

\begin{figure}[h!]
  \centering
  \includegraphics[width=85mm]{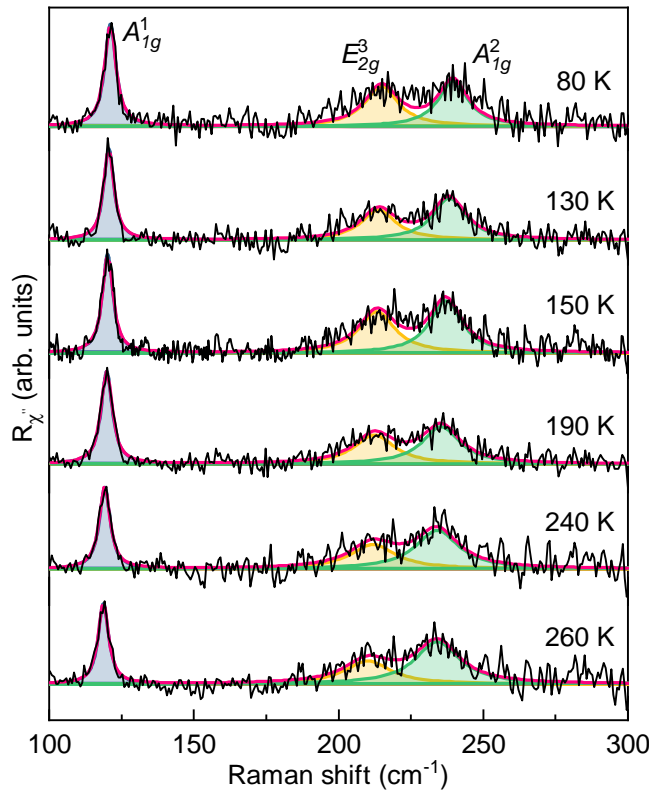}
  \caption{(Color  online) Modeled Raman spectra of \FGT single crystal, after subtracting continuum contributions, obtained at various temperatures. For experimental data modelling, Voigt line-shape was used.}
 \label{fig:FigureA1}
\end{figure}

In order to obtain temperature dependence of energies and line widths of observed \FGT phonon modes, Raman continuum, shown in coloured lines in Fig.~\ref{fig:Figure3}\,(a), was subtracted for simplicity from the raw Raman susceptibility data (black line). Spectra obtained after subtraction procedure are presented in Fig.~\ref{fig:FigureA1} (black line) for various temperatures. Because of the finite resolution of the spectrometer and the fact that line shapes of all the observed phonons are symmetric, Voigt line shape ($\Gamma_G= 0.8$ \wn) was used for data modelling. Blue, yellow and green lines in Fig.~\ref{fig:FigureA1} represent  fitting curves for $\Alg^1$, $\EZg^3$ and $\Alg^2$ phonon modes, respectively, whereas overall spectral shape is shown in red line.

\section*{Appendix C: Experimental details}
\label{sec:Exp_details}

Before being placed in a vacuum and cleaved, sample was glued to a copper plate with GE varnish in order to achieve good thermal conductivity and prevent strain effects. Silver paste, as a  material with high thermal conductivity, was used to attach the copper plate with the sample to the cryostat. Laser beam spot, focused through the Olympus long range objective of $\times$50 magnification, was approximately 6 $\mu$m in size, with a power less than 1 mW at the sample surface. 
TriVista 557 triple spectrometer was used in the subtractive mode, with diffraction grating combination of 1800/1800/2400 groves/mm and the entrance and second intermediate slit set to 80 $\mu$m, in order to enhance stray light rejection and attain good resolution.

\section*{Appendix D: Calculations}
\label{sec:Calc_comparison}

\begin{table}[h!]
\caption{Top panel: Comparison of calculated energies of Raman active phonons using relaxed (R) and experimental (non-relaxed - NR) lattice parameters for magnetic (M) and non-magnetic phase (NM), given in \wn. Obtained experimental values in magnetic phase at temperature of 80\,K are given in the last column. Bottom panel: Comparison of calculated energies of infrared optical phonons of \FGT.}
\label{ref:TableA1}
\begin{ruledtabular}
\centering
\resizebox{\linewidth}{!}{%
\begin{tabular}{c c c c c c}
\multicolumn{6}{c} {Raman active modes} \\

\cline{1-6} \\[-0.5em]

& \multicolumn{4}{c} {Calculations} & Experiment (M) \\ [1mm] \cline{2-5} \cline{6-6}  \\[-0.3em]
Sym. & NM-R & M-R & NM-NR & M-NR &\\ [1mm] \cline{1-6} \\[-0.5em]

$\EZg^1$ & 28.4	& 49.6 & 33.9 & 50.2& - \\[1mm]
$\Elg^1$ & 79.2  & 70.2 & 71.7 & 70.3 & - \\[1mm]
$\EZg^2$ & 115.5 & 121.0 & 100.0 & 122.2 & 89.2 \\[1mm]
$\Alg^1$ & 151.7 & 139.2 & 131.7 & 137.2 & 121.1 \\[1mm]
$\Elg^2$ & 225.5 & 206.0 &  194.3 & 209.5 & - \\[1mm]
$\EZg^3$ & 238.0 & 232.6 & 204.9 & 228.6 & 214.8 \\[1mm]
$\Alg^2$ & 272.0 & 262.6  & 235.7 & 233.4 & 239.6 \\[1mm]
$\EZg^4$ & 362.0 & 337.6 & 315.4 & 334.7 & - \\[1mm]
\hline \\[-0.5em]

\multicolumn{6}{c} {Infrared active modes} \\[1mm] \hline \\[-0.5em]

$\AZu^1$ & 70.7 & 96.6 & 73.5 & 92.7 & - \\[1mm]
$\Elu^1$ & 112.5 & 121.2 & 89.4 & 121.6 & - \\[1mm]
$\AZu^2$ & 206.0 & 162.5 & 183.1 & 153.7 &- \\[1mm]
$\Elu^2$ & 226.4 & 233.6 & 192.1 & 231.3 & - \\[1mm]
$\AZu^3$ & 271.8 & 248.6 & 240.8 & 241.0 & - \\[1mm]
$\Elu^3$ & 361.1 & 336.6 & 314.7 & 334.7 & - \\[1mm]
\end{tabular}}
\end{ruledtabular}
\end{table}

In the table \ref{ref:TableA1} results of DFT calculations are presented for magnetic (M) and non-magnetic (NM) relaxed, and experimental lattice parameters. For comparison, experimental results are shown in the last column. Since lattice parameters strongly depend on the Fe atoms deficiency, the best agreement with experimental results gives the magnetic non-relaxed solution.


\end{appendix}

\end{document}